Do altmetrics assess societal impact in a comparable way to case studies?

An empirical test of the convergent validity of altmetrics based on data from the

UK Research Excellence Framework (REF)[1]

Lutz Bornmann*, Robin Haunschild**, & Jonathan Adams***

*Division for Science and Innovation Studies
Administrative Headquarters of the Max Planck Society
Hofgartenstr. 8,
80539 Munich, Germany.
Email: bornmann@gv.mpg.de

**Max Planck Institute for Solid State Research
Heisenbergstr. 1,
70569 Stuttgart, Germany.
Email: R.Haunschild@fkf.mpg.de

*** ISI Clarivate Analytics, and,
The Policy Institute at King's,
King's College London,
22 Kingsway,
London WC2B 6LE, UK.
Email: jonathan.adams@kcl.ac.uk

---

[1] This paper is based on a manuscript which was presented at the Science & Technology Indicators (STI) conference in Leiden, the Netherlands (Bornmann, Haunschild, & Adams, 2018).

**Abstract**

Altmetrics have been proposed as a way to assess the societal impact of research. Although altmetrics are already in use as impact or attention metrics in different contexts, it is still not clear whether they really capture or reflect societal impact. This study is based on altmetrics, citation counts, research output and case study data from the UK Research Excellence Framework (REF), and peers' REF assessments of research output and societal impact. We investigated the convergent validity of altmetrics by using two REF datasets: publications submitted as research output (PRO) to the REF and publications referenced in case studies (PCS). Case studies, which are intended to demonstrate societal impact, should cite the most relevant research papers. We used the MHq' indicator for assessing impact – an indicator which has been introduced for count data with many zeros. The results of the first part of the analysis show that news media as well as mentions on Facebook, in blogs, in Wikipedia, and in policy-related documents have higher MHq' values for PCS than for PRO. Thus, the altmetric indicators seem to have convergent validity for these data. In the second part of the analysis, altmetrics have been correlated with REF reviewers' average scores on PCS. The negative or close to zero correlations question the convergent validity of altmetrics in that context. We suggest that they may capture a different aspect of societal impact (which can be called unknown attention) to that seen by reviewers (who are interested in the causal link between research and action in society).

# 1 Introduction

In his epochal report "Science, the Endless Frontier", Bush (1945) argues that publicly funded research would always (i.e. 'naturally') pay off for the society: excellent research will be followed by practical applications (LERU, 2017). The departure from the Bush philosophy became visible especially from the 1990s where governments expected objective evidence for the significance or excellence of academic science (Mostert, Ellenbroek, Meijer, van Ark, & Klasen, 2010). Performance-based research funding systems started with the Research Assessment Exercise (RAE) in the UK, followed by similar systems in more than ten countries worldwide (including New Zealand and Australia) (Ovseiko, Oancea, & Buchan, 2012). A UK Department for Business; Energy; Industrial Strategy (2016) overview shows that multiple national systems now also assess non-academic outputs and socio-economic outcomes and impacts.

In the European Union (EU), the explicit development of frameworks to measure (economic) returns from research started in the 2000s (Miettinen, Tuunainen, & Esko, 2015). Today, there is increasing demand from governments and funders of research that "researchers track the impact of their research to justify research expenditure by showing economic benefits, policy uptake, improved health and community outcomes, industry application and/or positive environmental effects" (Alla, Hall, Whiteford, Head, & Meurk, 2017). According to Derrick and Samuel (2016) the movement to societal impact can be described as a Kuhnian revolution for evaluation criteria (see also Bornmann, 2014a, 2016).

Prior to the current focus on societal impact, most 'impact' assessment focused on impact within the (academic) research environment as measured by counting citations to earlier academic publications by later academic publications. The broadening of the impact concept to include the economy and society led to a reconsideration of its definition. For example, overviews of the definition of and research on societal impact in general can be found in Bornmann (2013) and in De Silva and K. Vance (2017). Another definition, which we highlight, has been formulated on the basis of an overview of the societal impact literature and is very broad in its nature: "Research has a societal impact when auditable or recorded influence is achieved upon non-academic organisation(s) or actor(s) in a sector outside the university sector itself – for instance, by being used by one or more business corporations, government bodies, civil society organisations, media or specialist/professional media organisations or in public debate. As is the case with academic impacts, societal impacts need to be demonstrated rather than assumed. Evidence of external impacts can take the form of references to, citations of or discussion of a person, their work or research results" (Wilsdon et al., 2015, p. 6).

It is generally agreed that academic impact is reflected in citation analysis, whereas "measuring societal impact is problematic" (Moed, 2017, p. 7). Although alternative metrics (abbreviated as altmetrics) have been proposed as an indicator of societal impact, it is unclear whether such metrics are of practical use and value. In this study, using data from the UK Research Excellence Framework (REF) and from Altmetric (see www.altmetric.com), we investigate whether or not altmetric data can indeed provide an informative indicator in the context of societal impact. For each university REF submission in each subject-based Unit of Assessment (UoA, which conforms to a broad discipline), we compare the altmetric scores with traditional citation indicators and we compare grade point averages (GPAs) of case studies concerning societal impact with various altmetrics scores.

## 2    Altmetrics

Several definitions of altmetrics have been proposed; an overview can be found in Erdt, Nagarajan, Sin, and Theng (2016). A general definition, which subsumes many others, is that: "Altmetrics are non-traditional metrics that cover not just citation counts but also downloads, social media shares and other measures of impact of research outputs. The term is variously used to mean 'alternative metrics' or 'article level metrics', and it encompasses webometrics, or cybermetrics, which measure the features and relationships of online items, such as websites and log files. The rise of new social media has created an additional stream of work under the label altmetrics. These are indicators derived from social websites, such as Twitter, Academia.edu, Mendeley, and ResearchGate with data that can be gathered automatically by computer programs" (Wilsdon et al., 2015, pp. 5-6). Moed (2017) and others have suggested that altmetrics primarily reflect 'attention' rather than impact or influence (see www.altmetric.com).

According to Adie (2014b), three developments led to the use of altmetrics for measuring research impact: (1) the wish to measure a return on investments; (2) the shift from print to online for documenting and publishing research; and (3) the publication of the altmetrics manifesto (Priem, Taraborelli, Groth, & Neylon, 2010), which gave the area a clear focal point and label. Several journals have recently published special issues on research around altmetrics (e.g., *Aslib Journal of Information Management* and *Bulletin of the Association for Information Science & Technology*) (Das & Mishra, 2014). Although such studies are at an early stage (Zahedi, Costas, & Wouters, 2014), research overviews exist (Galloway, Pease, & Rauh, 2013; Sugimoto, Work, Larivière, & Haustein, 2017; Torres-Salinas, Cabezas-Clavijo, & Jimenez-Contreras, 2013). González-Valiente, Pacheco-Mendoza, and Arencibia-Jorge (2016) found 253 relevant publications between 2005 and 2015. A topical overview (e.g., user group differences and limitations of altmetrics) is given by Erdt et al. (2016).

There is diversity in the range of data sources that altmetrics can link to research publications (e.g., views, downloads, clicks, notes, saves, tweets, shares, likes, recommends, tags, posts, trackbacks, discussions, bookmarks, and comments) and these can be grouped in multiple ways. For example, Moed (2017) identified four categories of sources: (1) social media such as Twitter and Facebook, covering social activity; (2) reference managers or reader libraries such as Mendeley or ResearchGate, covering scholarly activity; (3) variants of scholarly blogs, reflecting scholarly commentary; and (4) mass media coverage such as newspapers and news broadcasts, informing a wider public. Sugimoto et al. (2017) and Haustein (2016) used six categories to aggregate the sources: "(a) social networking (e.g., Facebook, ResearchGate); (b) social bookmarking and reference management (e.g., Mendeley, Zotero); (c) social data sharing including sharing of datasets, software code, presentations, figures and videos, etc. (e.g., Figshare, Github); (d) blogging (e.g., ResearchBlogging, Wordpress); (e) microblogging (e.g., Twitter, Weibo); (f) wikis (e.g., Wikipedia); (g) social recommending, rating and reviewing (e.g., Reddit, F1000Prime )" (Haustein, 2016, p. 417).

      The potential use of altmetrics for measuring broader (i.e. non-academic) impact has been widely discussed. Adie and Roe (2013) speak about "measuring reach amongst audiences who do not traditionally cite" (p. 13). Altmetrics may capture the discourse around scientific contributions by people outside the core scientific community (NISO Alternative Assessment Metrics Project, 2014). Sugimoto et al. (2017) also note that altmetrics "has been advocated as a potential indicator of such [societal] impact" (p. 2046). The National Information Standards Organization (2016) suggest to stakeholders who are interested in the broad influence of scholarly outputs that the use of altmetrics "may offer insight into impact by calculating an output's reach, social relevance, and attention from a given community, which may include members of the public sphere" (p. 1). For Taylor (2013), altmetrics form an important element in the relationship between scholarly and social parties in society: "With all parties having an interest in impact (both scholarly and social) and reach (again, both scholarly and social), the promise of altmetrics is, at the very minimum, to provide some description of the reach of scholarly impact" (p. 30). According to Priem (2014) "an important property of altmetrics is the ability to track impact on broad or general audiences, as well as on scholars" (p. 267).

The range, categorization and use of altmetric variables as indicators of the non-traditional impact of academic research is of current interest because societal impact assessment was introduced into formal research assessment as part of the UK's REF in 2014, and the possibility of using altmetrics in this context has been suggested (Adie, 2014a). For Thelwall and Kousha (2015a), "in theory, alternative metrics may be helpful when evaluators, funders or even national research assessment need to know about 'social, economic and cultural benefits and impacts beyond academia' (REF, 2011, p. 4) as well as non-standard impacts inside academia" (p. 588). There is, consequently, significant policy interest in such options being tested.

## 3 Measuring societal impact in the UK Research Excellence Framework by using case studies

The REF is an important context for the study of altmetrics. REF2014 is the first national evaluation system to include societal impact criteria in the allocation of research funding, with such impact contributing up to 20% of the overall assessment that determines core university research funds over 5-8 years periods (Samuel & Derrick, 2015). Its predecessors have become global reference points for evaluation practice, so it will be influential. But REF evaluation is complex and costly: RAND Europe estimated that UK higher education institutions (HEIs) spent £55 million to demonstrate impact for the REF 2014 (Chowdhury, Koya, & Philipson, 2016); and total REF running costs were around £246 million – according to the UK Department for Business, Energy, and Industrial Strategy (2016).[2]

For the REF, societal impact was defined as "an effect on, change or benefit to the economy, society, culture, public policy or services, health, the environment or quality of life, beyond academia" (see www.hefce.ac.uk/rsrch/REFimpact). The research producing the impact was expected to benefit a specific sector in society and to be of a quality that matched a reasonably high international standard (Terama, Smallman, Lock, & Johnson, 2017). Evidence of this impact could have various forms, for example: "references to, citations of or discussion of an academic or their work; in a practitioner or commercial document; in media or specialist media outlets; in the records of meetings, conferences, seminars, working groups and other interchanges" (Wilsdon et al., 2015, pp. 44-45).

---

[2] This is the total cost. The UK REF can also be seen as an efficient way of distributing competitive performance related research funds across all submitting institutions. The transaction costs (i.e. total costs as % total funding distributed) are around 3%. This is significantly less than the >10% cost of delivering UK Research Council project grants, which does not account for the costs of unsuccessful applicants.

It should be understood that there were two distinct areas of REF assessment: research achievement, assessed via outputs; and research impact, assessed via case studies. For research outputs, institutions submitted up to four items (published over the census period since the last RAE in 2008) for each assessable researcher. Each research output was read by a subject-based peer panel and graded on a 5-point scale (0*-4*), where 0* indicated no research content and 4* indicated 'world leading quality" (see https://www.ref.ac.uk/2014/pubs/refmanagersreport/). For the assessment of societal impact, four-page case studies were used and each case study covered the work (over a period up to 20 years) of a group of several researchers. The case studies were graded qualitatively by the same subject-based panel using the same reference scale as research output. Case studies followed a standard template with a summary, description of underpinning research, research reference, description of the impact, and corroborative sources. All case studies were indexed and are available in a report and website (Digital Science, 2016).

The challenge for HEIs is to select and describe research with quality (in terms of originality, significance, and rigor) and where research impact on other sectors can be evidenced (Samuel & Derrick, 2015). The case study method for doing this has advantages and disadvantages. One key advantage is that the complex links between research and impact can be explained in an adequate manner that satisfies governmental expectations, as it did in this instance: case studies "provide a rich picture of the variety and quality of the contribution that UK research has made across our society and economy. The resulting database of case studies is a unique and valuable source of information on the impact of UK research" (Department for Business; Energy; Industrial Strategy, 2016, p. 21).

Disadvantages of the case study approach also appear (van Noorden, 2015). First, each case study is unique, reducing comparability between HEIs. The results cannot be generalized from the research explained in the case study to a unit at a higher level. Khazragui and Hudson (2015) found little evidence "for a homogenized … version of impact dominating the REF submissions" (p. 60). Second, case studies tend to be biased "towards 'good news' stories" (Raftery, Hanney, Greenhalgh, Glover, & Blatch-Jones, 2016, p. xxiii). Thus, they do not give insights on the overall return on research investment (Khazragui & Hudson, 2015). Third, case studies are expensive to write (see footnote 2): University College London alone "wrote 300 case studies that took around 15 person-years of work, and hired four full-time staff members to help, says David Price, the university's vice-provost for research" (van Noorden, 2015, p. 150).

Rich data on diverse impacts are rarely delivered through sets of standardized quantitative indicators. Raftery et al. (2016) therefore warned of the danger "of excluding case studies in favour of metrics in future such exercises" (p. 79). However, the disadvantages of case studies have stimulated the search for indicators that could potentially replace, or at least supplement, this costly approach; for example, "a more consistent toolkit of impact metrics that can be more easily compared across and between cases" (Wilsdon et al., 2015, p. 49). In this context, the possible application of altmetrics has been frequently mentioned and thus needs to be tested.

# 4 The UK research evaluation process as a target of empirical research

Since many datasets were made available from the REF and previous RAEs (see, e.g., http://impact.ref.ac.uk/CaseStudies), a number of analytical studies have targeted this accessible example of a national system. For example, Wooding, Van Leeuwen, Parks, Kapur, and Grant (2015) found significant discrepancies in measurements of quality improvements between two REF periods: the improvements indicated by comparison of peer assessments appear to be higher than the improvements indicated by bibliometric data.

Many studies have investigated the correlation between REF output scores and traditional metrics, but it must be understood that publicly available data do not reveal scores for individual documents or case studies. Scores are in fact aggregated at unit level. Figure 1 shows the results of a meta-analysis (Glass, 1976) based on the correlation coefficients reported in previous studies. We calculated a random effects model which considers that the single results are based on somewhat different populations (Cumming & Calin-Jageman, 2016). If publications report more than one coefficient, all coefficients have been included in the meta-analysis. However, the clustered nature of the data has been considered by including a moderator variable representing the single publications in the model.

Across the nine studies (including 40 correlation coefficients), the random effects meta-analysis indicates, with $r_{pooled} = 0.71$, a correlation between REF scores and citations, which can be interpreted approximately as a large effect (Cohen, 1988). With 95% confidence interval (CI) [0.64, 0.78], the CI is fairly small. This is entirely unsurprising given the nature of the exercise: it would be astounding if (normalized) citation scores did not broadly correlate with REF outcomes. The correlations in Figure 1 have, however, been calculated across different contexts using variables affected by metrics, subject areas, citation indexes, and time periods. Context differs not only between the studies, but also within the studies, so the combined set is heterogeneous and challenging. All correlations are Spearman rank order correlations which are based either on the scores from the REF (or the former Research Assessment Exercise) or ranking positions based on those scores.

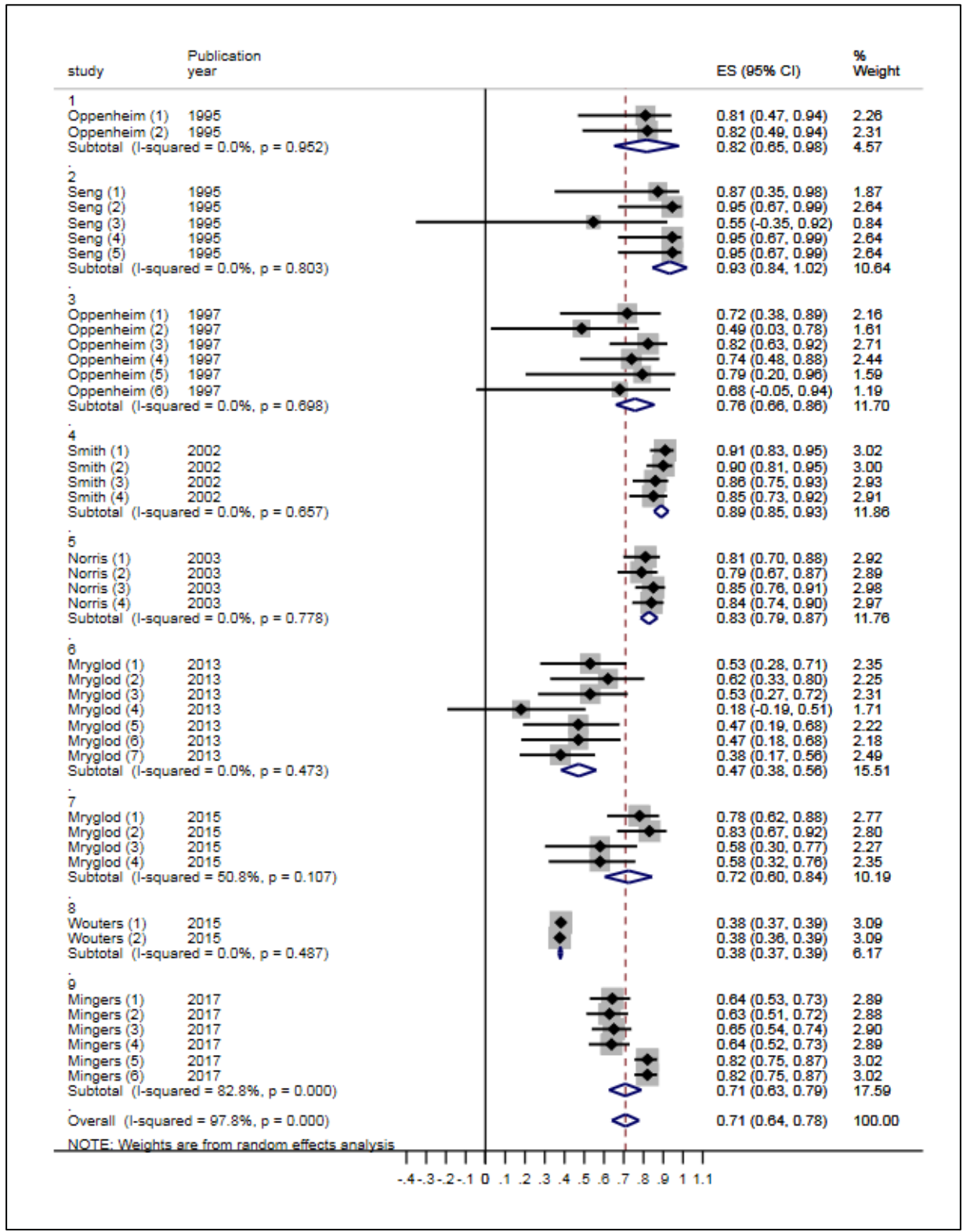

Figure 1. Results of a random effects meta-analysis of nine studies reporting 40 coefficients for the correlation between REF (or the former Research Assessment Exercise) output scores and bibliometrics

In addition to the studies comparing REF scores and bibliometrics, other studies have focused on case studies and on altmetrics. In the following, we describe in more detail those studies that are most relevant for the present study:

(1) In the REF2014 study by HEFCE (2015), published alongside the "Metric Tide" (a report of the independent review of the role of metrics in research assessment and management; Wilsdon et al., 2015), attention indicators were correlated with REF output (articles, books, and other outputs) and quality scores (0* to 4*) at the level of UOA. Correlations including a range of traditional and alternative metrics were computed on this level. The results are as follows: "Percentile metrics [numbers of highly cited publications] were significant for the Physics UOA outputs, while citation and index measures such as citation count, SJR [SCImago Journal Rank], FWCI [field-weighted citation impact] and SNIP [source normalized impact per paper] were significant for the Earth Systems and Environmental Sciences UOA. Altmetrics such as Mendeley readership, Google Scholar and tweets were all significant for Biological Sciences outputs, suggesting that these UOAs correlated well with a number of metrics measures" (HEFCE, 2015, p. 14).

(2) The study by Ravenscroft, Liakata, Clare, and Duma (2017) used a similar approach to that in this study. They focused on the references cited in case studies and correlated the average altmetric scores for the references with the average REF scores concerning societal impact. They used the Altmetric API and were able to append the Altmetric Attention Score (weighted count including different altmetrics, e.g., tweets and blog mentions) to the publications referenced in case studies for only around 60% of the publications. Pearson correlation coefficients were close to zero and the negative coefficient ($r = -0.0803$) suggests that the scores reflect different constructs. Thus, their work does not support the convergent validity of altmetrics with another indicator measuring societal impact.

(3) The study published by Digital Science (2016) included all publications that had been submitted to REF for assessment between 1988 and 2013. The study is based on submitted outputs and case study references of REF 2014. The study followed an initial assessment of the nature, scale and beneficiaries of the UK universities' research impact (King's College London and Digital Science, 2015) which is based on an analysis of the case studies themselves (submitted to the 2014 REF). In the new report, Digital Science (2016) was especially interested in "the relationship between submitted outputs (as evidence of academic impact) and case study references (as evidence of societal and economic impact)" (p. 4). The results show that a consistent 42% of the referenced publications drawn from each publication year had also been submitted as outputs in previous RAEs. Around 60% of these referenced publications were published in the same census period as REF 2014; the rest were published in earlier RAE census cycles. The references in case studies focused on recent research and the overlap between submitted outputs and referenced publications was lower in basic research areas than in applied research areas. The authors concluded that "work of fundamental quality regularly leads to a wider range of benefits for wealth creation and the quality of life" (Digital Science, 2016, p. 19).

(4) Morrow, Goreham, and Ross (2017) investigated a subsample of 1,309 research-based case studies in leadership, governance, and management. They used a mixed-method approach to analyze the case studies. The analysis reveals, for example, the following prevalent types of impact reported: "government policy (n = 676, 52% of cases), training (n = 615, 47% of cases), impact on understanding (n = 506, 39% of cases) and strategy (n = 478, 37% of cases)" (Morrow et al., 2017, p. 419).

# 5 Convergent validity of altmetrics data for measuring societal impact

According to the US National Research Council (2014) "no high-quality metrics for measuring societal impact currently exist that are adequate for evaluating the impacts of federally funded research on a national scale" (p. 70). Altmetrics has been proposed as a possible candidate for quantitative measurement of societal impact. Systematic evidence to support this does not exist yet (Haustein, Bowman, et al., 2014), but altmetrics are: "being (too) quickly identified and used as indicators of impact and scientific performance" (Haustein, 2016, p. 414). Thelwall and Kousha (2015b) conclude that "of all the indicators reviewed … only Google Patents citations and clinical guideline citations clearly reflect wider societal impact and no social media metrics do" (p. 615). The most important problem is that it is not clear what conclusions can be drawn "when an article is frequently mentioned within the social web" (Barthel, Tönnies, Köhncke, Siehndel, & Balke, 2015, p. 119). The meaning of altmetrics thus remains obscure (Taylor & Plume, 2014; Zahedi et al., 2014).

Large-scale, systematic studies of altmetrics are necessary to reveal evidence of their validity as an impact indicator (Haustein, Peters, Sugimoto, Thelwall, & Larivière, 2014). In this study, (following earlier studies by Bornmann, 2014b, 2015b; Ravenscroft et al., 2017), we address the question of convergent validity of altmetrics data for indexing societal impact by:

(1) Comparing the impact of two groups of publications:

    a. PCS = publications cited in case studies: the literature included in case study 'References' should support the underpinning research behind the reported impact; we therefore expect high societal impact for the cited publications;

  b. PRO = publications submitted as REF Outputs: these evidence an individual's highest quality of research; we therefore expect lower societal but higher citation impact.

(2) Determining, as a test of convergent validity, whether two different approaches of measuring societal impact (REF scores on case studies and altmetrics) create comparable measures of a common construct (societal impact) (Picardi & Masick, 2013; Thorngate, Dawes, & Foddy, 2009).

(3) Studying discriminant validity by analyzing societal and citation impact of PCS and PRO.

(4) Further considering the REF scores for comparing altmetrics with expert evaluation of societal impact (as suggested, e.g., by Priem et al., 2010): if both measure the same construct and are convergently valid, then the REF impact scores should substantially correlate with altmetrics.

# 6 Reflections about measuring societal impact and the possible role of altmetrics

There are a number of schema used in evaluation contexts for categorising types of societal impact. These include the widely used PEST typology (political, economic, social, technological); which may extend to PESTLE (add legal and environmental); which was extended by REF impact analysis (see https://impact.ref.ac.uk/casestudies/) to PESTLE-HC (add health and cultural). A further variant is (social, cultural, environmental, economic) as returns to society from publicly funded research (Donovan, 2011). However, while many governments are interested in describing and measuring societal impact, there is "a lack of clearly documented, empirical … research impact evaluations" (Bell, Shaw, & Boaz, 2011, p. 227) (e.g., Lebel & McLean, 2018). Because a generally accepted framework, based on adequate datasets and specific methods for measuring societal impact, does not exist, the focus of most quasi-quantitative societal impact is primarily on economic impact using established R&D indicators (Godin & Dore, 2005).

According to Martin (2007), societal impact measurements have four problems (see also Bornmann, 2012, 2013): (1) <u>Causality</u>: It is not clear what cause determines what impact and what impact results from what cause. (2) <u>Attribution</u>: Societal impact can be complex, diffuse, and contingent; it is scarcely determinable which role certain pieces of research play. (3) <u>Internationality</u>: Research is an international endeavour including many actors so unique national (or institutional) contributions are difficult to determine. (4) <u>Evaluation timescale</u>: Research evaluations focus on recent years but societal research benefits lag by many years so research benefits lie beyond the evaluation process.

Three further problems are as follows: (5) Field-specific differences: Research outcomes from technology, sociology and cultural disciplines have very distinct impacts affecting specific sectors. It is therefore infeasible to develop a unique and common assessment mechanism akin to citation analysis for academic impact. (6) Institutional heterogeneity: There is no unique model of a successful university and foci vary between teaching and research and between cultural and national priorities, so the informed matching of impact outcomes and objectives is localised and complex. (7) Positive and negative impacts: Societal impact might be positive and/ or negative. "Environmental research that leads to the closure of a fishery might have an immediate negative economic impact, even though … it will preserve a resource that might again become available for use. The fishing industry and conservationists might have very different views …" (Rymer, 2011, p. 6). Impact measurements would need to be complex to account for this.

# 7     Methods

## 7.1     Description of altmetrics

In this study, we included six sources for altmetric indicators that are frequently used in altmetrics studies (an overview of these sources can be found in Thelwall (2017)).

**Twitter** (see www.twitter.com) is a popular microblogging platform (founded in 2006). Tweets may contain references to scientific papers. According to Haustein, Peters, Bar-Ilan, et al. (2014) "Twitter activity reflects discussion around these articles" (p. 1148). The study by Yu (2017) suggested that these discussions are led by public users. Andersen and Haustein (2015) investigated the hypothesis that medical research, which is close to clinical practice, is more popular for Twitter users than basic research. Since the results confirm the hypothesis, tweets might reflect attractiveness of papers for a broader audience. However, the results of Robinson-Garcia, Costas, Isett, Melkers, and Hicks (2017) point out the problematic nature of Twitter data for research evaluation: "A multi-year campaign has sought to convince us that counting the number of tweets about papers has value. Yet, reading tweets about dental journal articles suggested the opposite. This analysis found: obsessive single issue tweeting, duplicate tweeting from many accounts presumably under centralized professional management, bots, and much presumably human tweeting duplicative, almost entirely mechanical and devoid of original thought". Many tweets merely mention the title of an article or its brief summary (Thelwall & Kousha, 2015b). In this study, the number of tweets (including retweets) with some reference (detected by Altmetric) to a scientific paper in our dataset is counted.

**Wikipedia** (see https://www.wikipedia.org) is a free encyclopedia platform with editable content (Mas-Bleda & Thelwall, 2016). Most of the references in Wikipedia are to non-scholarly documents but contributors include scholarly references (Priem, 2014). Serrano-López, Ingwersen, and Sanz-Casado (2017) undertook a Wikipedia case study with papers on Wind Power and found that < 1% of relevant papers have been cited on Wikipedia "implying that the direct societal impact through the Wikipedia is extremely small for Wind Power research" (p. 1471). Based on their results, the authors do not recommend Wikipedia data for research evaluation. Kousha and Thelwall (2017) used a significantly larger sample of papers in their study than Serrano-López et al. (2017) and found that only 5% had any citation from Wikipedia. Further studies on the use of Wikipedia as source of altmetrics can be found in Sugimoto, Work, Larivière, and Haustein (2016) and Serrano-López et al. (2017). In this study, the number of Wikipedia articles with some reference (detected by Altmetric) to a scientific paper in our dataset is counted.

**Policy-related documents** have only recently been analyzed in an altmetrics context. Mentions are discovered by text mining databases of, e.g. World Health Organization or European Food Safety Authority (Bornmann, Haunschild, & Marx, 2016; Haunschild & Bornmann, 2017): "The influence on policy is an important asset associated to the notion of broader societal influence" (OPENing UP, 2016, p. 24). However, the results of Haunschild and Bornmann (2017) show that "less than 0.5% of the papers published in different subject categories are mentioned at least once in policy-related documents" (p. 1209). They also reported that Altmetric tracks more than 100 policy sources as of December 19, 2015. Analyzing a rather large set of 191,276 publications from the policy-relevant field of climate change, Bornmann et al. (2016) found that "only 1.2 % (n = 2341) have at least one policy mention" (p. 1477). In this study, the number of policy-related documents with some reference (detected by Altmetric) to a scientific paper in our dataset is counted.

**Blogs** are one of the earliest social media platforms (Bik & Goldstein, 2013). Blogs are written about scholarly papers. Citations of publications in a formal or informal way (Shema, 2014) can be collated, although informal citing leads to uncertainty (Shema, Bar-Ilan, & Thelwall, 2014). Blogging has been seen as a bridge between research and the public (Bar-Ilan, Shema, & Thelwall, 2014; Bonetta, 2007). Mewburn and Thomson (2013) show, however, that "this was one of the less popular motivations for academics to blog" (p. 1113). An overview of Blog studies has been published by Work, Haustein, Bowman, and Larivière (2015). In this study, the number of blog posts with some reference (detected by Altmetric) to a scientific paper in our dataset is counted.

**Facebook** is a widely used social networking and social media platform (Bik & Goldstein, 2013). Users can share papers with other users, so mentions of papers in posts or Facebook likes of papers can be counted. Ringelhan, Wollersheim, and Welpe (2015) tested whether Facebook likes are an information indicator of scholarly impact. Their results indicate "an interdisciplinary difference in the predictive value of Facebook likes, according to which Facebook likes only predict citations in the psychological area but not in the non-psychological area of business or in the field of life sciences". In this study, the number of Facebook posts with some reference (detected by Altmetric) to a scientific paper in our dataset is counted. Likes were not included.

**News** attention (e.g., by the *New York Times*) refers to scientific papers mentioned (via direct links or unique identifiers) in news reports (Priem, 2014). Thus, the attention received by the papers can be counted. We found more than 2,000 different news sources in our altmetrics dataset from November 2017. In this study, the number of news articles with some reference (detected by Altmetric) to a scientific paper in our dataset is counted.

### 7.2  Mantel-Haenszel quotient (MHq')

Bornmann and Haunschild (2018) proposed the MHq' indicator as a field- and time-normalized altmetrics indicator (see also Haunschild & Bornmann, 2018), because it is especially designed for count-data with many zeros, which is typical of most altmetrics data and means that the usual bibliometric normalization procedures cannot be applied (Haunschild, Schier, & Bornmann, 2016). The empirical results of Bornmann and Haunschild (2018) point out that the indicator has convergent validity, because, on the basis of citation data, it distinguishes between different quality levels in terms of peers' assessments. The following explanation of the MHq' indicator is based on Bornmann and Haunschild (2018) and Haunschild and Bornmann (2018).

In contrast to many other normalized indicators in bibliometrics, MHq' is not calculated on the single paper level, but on an aggregated level considering field and time of publication. For the impact comparison of publication sets (here: PRO and PCS) with reference sets, the 2×2 cross tables (which are pooled) consist of the number of papers mentioned and those not mentioned in each combination ($f$) of subject category and publication year. Thus, in the 2×2 subject- and publication year-specific cross table with the cells $a_f$, $b_f$, $c_f$, and $d_f$ (see Table 1):

$a_f$ is the number of mentioned papers in set $g$ (papers of a UK university = UKPRN in data coding) in subject category and publication combination year $f$;

$b_f$ is the number of not mentioned papers in set $g$ (papers of a UKPRN) in subject category and publication year combination $f$;

$c_f$ is the number of mentioned papers in subject category and publication year combination $f$;

$d_f$ is the number of not mentioned papers published in subject category and publication year combination $f$.

Since MHq' compares groups of papers, the papers of set $g$ are not part of the papers in the 'world' comparator set (here = all papers submitted to the REF).

Table 1. 2 x 2 subject- and publication year-specific cross table

|         | Number of mentioned papers | Number of not mentioned papers |
|---------|----------------------------|--------------------------------|
| Group $g$ | $a_f$ | $b_f$ |
| World   | $c_f' = c_f - a_f$ | $d_f' = d_f - b_f$ |

We start by defining some dummy variables for the MH analysis:

$$R_f = \frac{a_f d_{f'}}{n_f} \text{ and } R = \sum_{f=1}^{F} R_f, \quad (1)$$

$$S_f = \frac{b_f c_{f'}}{n_f} \text{ and } S = \sum_{f=1}^{F} S_f, \quad (2)$$

$$P_f = \frac{a_f + d_{f'}}{n_f} \text{ and } Q_f = 1 - P_f \quad (3)$$

Where $n_f = a_f + b_f + c_{f'} + d_{f'}$

MHq' is simply:

$$\text{MHq}' = \frac{R}{S} \quad (4)$$

The confidence intervals (CI) for MHq' are calculated following Fleiss, Levin, and Paik (2003) and the variance of ln MHq' is estimated by:

$$\widehat{Var}(\ln MHq') = \frac{1}{2} \left\{ \frac{\sum_{f=1}^{F} P_f R_f}{R^2} + \frac{\sum_{f=1}^{F} (P_f S_f + Q_f R_f)}{RS} + \frac{\sum_{f=1}^{F} Q_f S_f}{S^2} \right\} \quad (5)$$

The confidence interval for the MHq' can be constructed with

$$MHq_L' = \exp\left[\ln(MHq') - 1.96\sqrt{\widehat{Var}[\ln(MHq')]}\right] \quad (6)$$
$$MHq_U' = \exp\left[\ln(MHq') + 1.96\sqrt{\widehat{Var}[\ln(MHq')]}\right] \quad (7)$$

## 7.3 Dataset used

The REF output data were downloaded from http://results.ref.ac.uk/DownloadSubmissions/ByForm/REF2 on 28 September 2017. There were 250,043 papers submitted as REF outputs (PRO) of which 149,616 (59.8%) had a DOI. REF case study IDs and corresponding DOIs were shared with us by Digital Science (2016) on 11 December 2017. We used the REF application programming interface (API, http://impact.ref.ac.uk/CaseStudies/APIhelp.aspx) via R (R Core Team, 2014) to append the UoA and UKPRN to each REF case study ID. There were 6,642 case study IDs of which we excluded 15 (0.23%) where the API returned an error instead of UKPRN and UoA. Of the 36,244 papers referenced in case studies (PCS), 25,313 had a DOI. Papers submitted multiple times (as REF outputs; case study references; or as both) were retained in the data set because they were always associated with different submissions and hence different peer grades and scores. A total REF score was calculated as a geometrically weighted sum from the ratings (score = $(4 \times 4^*) + (3 \times 3^*) + (2 \times 2^*) + (1 \times 1^*)$) and an average was calculated for each UKPRN.

Bibliometric data from Elsevier's Scopus database (see https://www.scopus.com) were used via a custom database of the Competence Center for Bibliometrics (see http://www.bibliometrie.info/), last updated in April, 2017. All papers submitted as PRO or PCS were matched via their DOI with the Scopus database and the number of citations and a Scopus subject area (see https://service.elsevier.com/app/answers/detail/a_id/15181/supporthub/scopus/) were appended to each DOI. Of the PCS case study references, 17,525 (69.2%) could be found in the Scopus database via their DOI and 126,694 (84.7%) of the PRO REF outputs could be similarly matched. Citations were determined for all papers published before 2015 using a two-year citation window, chosen as a compromise that avoided a sparse dataset but kept a relationship to longer citation windows (Glänzel & Schöpflin, 1995). The papers published in 2015 and 2016 (n=49 papers) were excluded, first, because the citation window was unduly short and, second, because they were not part of the REF census period. The Scopus subject areas were aggregated to a high level where ABCD subject codes were merged, subsuming C and D levels into the subject code AB00. Some papers were assigned to multiple aggregated

Scopus subject areas. In this study, there are 732 overlapping Scopus subject areas, referred to as fields, constructed from the multiple classification (Rons, 2012, 2014).

Some papers are mentioned in multiple case studies and submitted multiple times as REF outputs by different UKPRN – UoA combinations. In total, 138,309 papers are included in our analysis of which: 11,822 belong to PCS but not PRO; 120,784 belong to PRO but not PCS; and 5,703 are common to both groups (recorded in PRO and PCS).

In order to check the representativeness of the refined dataset used for analysis, we calculated the average impact scores (i) within the full REF dataset and (ii) only for the UKPRN – UoA combinations which had at least one case study paper in our dataset. The results shown in Table **2** indicate that (i) our focused data seems to be representative of the full dataset and (ii) the average ratings are high: 3* and 4* ratings account for ~75%.

Table 2: Average REF ratings in the full and our considered dataset (row percentage)

| Dataset | Average 1* | Average 2* | Average 3* | Average 4* |
| --- | --- | --- | --- | --- |
| Full | 5.2 | 19.9 | 41.6 | 31.7 |
| Considered in our study | 5.3 | 19.5 | 41.6 | 32.4 |

Altmetrics data were sourced from a locally maintained database using data shared with us by Altmetric (02 October 2017). The data include altmetric counts from the following sources and areas (see Haustein, 2016): social networking (e.g., Facebook counts); professional networking (e.g., LinkedIn); blogging (e.g., ScienceBlogs); microblogging (e.g., Twitter counts); wikis (e.g., Wikipedia); and policy-relevant usage (e.g., policy documents). We appended a mention count to each DOI using the following altmetrics sources: Twitter, Facebook, blogs, news, policy documents, and Wikipedia. A DOI not known to the altmetrics database was recorded as 'not mentioned'.

# 8      Results

## 8.1      Analyses of metrics scores for three groups of papers

We compared altmetrics scores (e.g., tweets) with traditional citation scores for three groups of papers. Our expected metrics scores for the different groups (taking into account the required normalization for field and year) are shown in Table 3:

(1) Higher altmetrics scores for PCS than for PRO;

(2) Relatively higher citation scores for PRO than for PCS.

A significant number of PCS were also PRO in the REF or previous RAEs (Digital Science (2016) and might have shared or hybrid characteristics. We should predict the highest combined scores for the overlap group (PCS & PRO) because these papers would attract attention both for academic research impact and for socio-economic impact, with a potential interactive pushing effect. The data are therefore clustered and analyzed in three groups:

(1) PCS (not PRO): 11,822 papers

(2) PRO (not PCS): 120,784 papers

(3) PCS & PRO (common to both): 5,703 papers

Table 3. Analyzing convergent and discriminant validity in this study: expected metrics scores for three groups of papers

|  | **PCS (not PRO)** | **PRO (not PCS)** | **PCS & PRO (common to PCS and PRO)** |
|---|---|---|---|
| **Altmetrics** | Higher | Lower | Highest |
| **Citation impact** | Lower | Higher | Highest |

The results of the analyses are shown in Figure 2 which displays MHq' values for PCS, PRO, and [PCS&PRO] with upper and lower bounds of 95% CIs. For traditional citation metrics, the scores are in accordance with expectations: citation impact for PCS is significantly lower than that for PRO, while impact for [PCS&PRO] is similar to but somewhat higher than for PRO. For altmetrics, we have very different results, but ones that also conform to the expectations in Table 3.

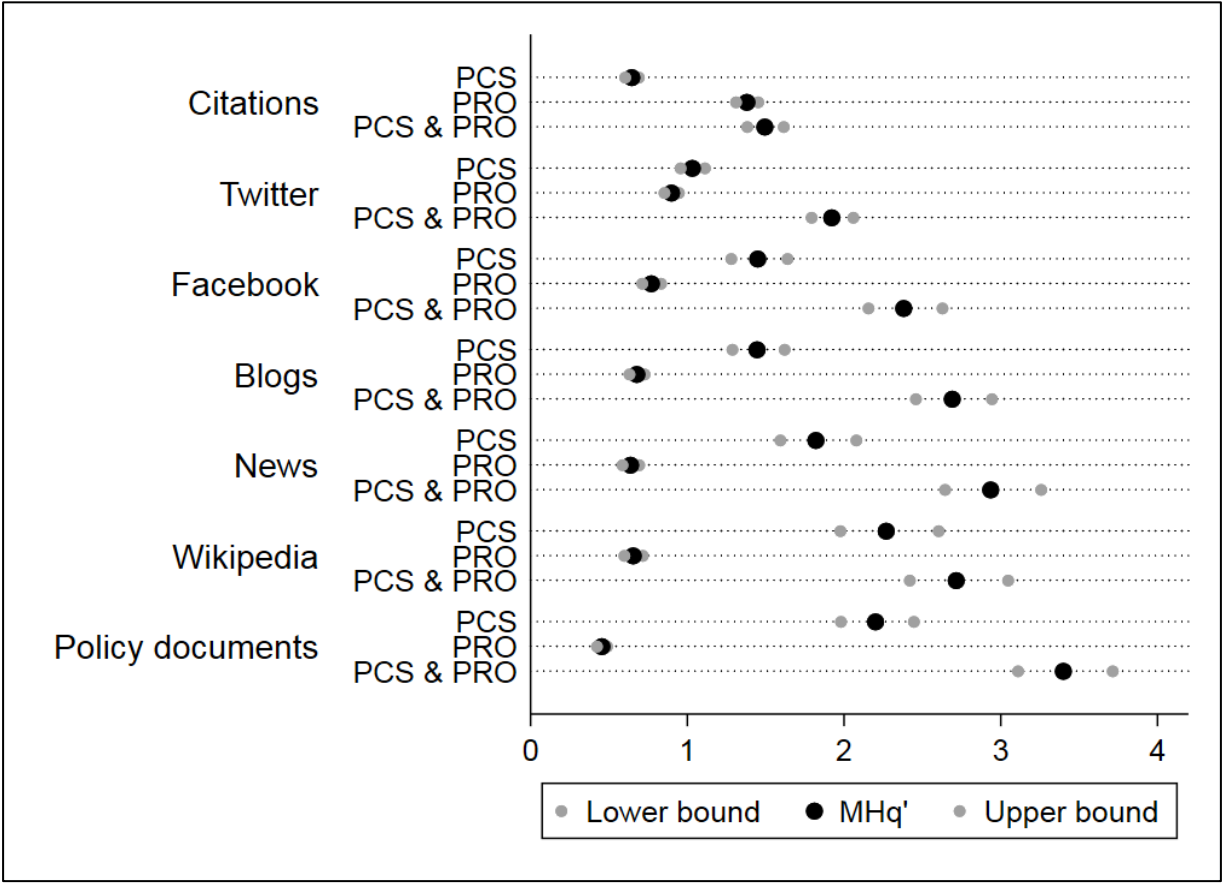

Figure 2. MHq' values for PCS, PRO, and [PCS & PRO] separated by different metrics (citations and altmetrics). The altmetrics are sorted by the impact differences between PCS and PRO.

In practice, the scores contributed by various altmetric sources differ in the degree of impact differences between PCS and PRO and are listed by these differences in Figure 2. Consistent results are visible for mentions of papers in policy-related documents and Wikipedia: the altmetrics impact of PCS is (significantly) higher than that of PRO (especially regarding mentions in policy-related documents). The result for papers mentioned in news items is similar, though slightly smaller, and is also statistically significant. Smaller impact differences between PCS and PRO are visible for blogs and Facebook.

The difference between PCS and PRO scores is very small: indeed, it is close to zero for the scores based on Twitter counts. These results might speak against their use for serious societal impact measurements. There is also a lack of correlation between tweets and citations (Bornmann, 2015a) so they do not appear to reflect impact on academic research. Twitter counts are widely noted and displayed as scores (e.g. on university web-pages) but the question remains as to what impact, if any, they actually reflect.

**8.2    Correlation between metrics scores and assessments by peers**

We considered REF scores for the output and impact dimensions: Table 4 shows the different comparisons and expected outcomes and Table 5 shows the varying numbers of institutional units that are available for analysis. The empirical results are presented in Figure 3. Since the REF scores are only available on the institutional level (and not on the level of single publications), PRO and PCS could not be separated into three groups, as was done in Table 3, but remain as two overlapping groups. Thus, various papers are considered multiple times. We expect higher correlations (i) for PRO between REF output scores and citation impact and (ii) for PCS between REF impact scores (for case studies) and altmetrics. Lower correlations are expected for (i) REF output scores and altmetrics for PRO and (ii) REF impact scores (for case studies) and citation impact for PCS.

Table 4. Analyzing convergent and discriminant validity: expected correlations between REF scores (for output and case studies) and metrics scores

|  |  | **Average REF scores/publication set** | |
|---|---|---|---|
|  |  | **Output scores for PRO** | **Impact scores for PCS** |
| **Metric scores** | **Altmetrics** | Lower | Higher |
|  | **Citation impact** | Higher | Lower |

Table 5. Number of units (UKPRN) considered in the analysis

|  | **Average REF scores/publication set** |
|---|---|

|  | **Output scores for PRO** | **Impact scores for PCS** |
|---|---|---|
| **Citations** | 141 | 106 |
| **Twitter** | 135 | 110 |
| **Policy documents** | 122 | 87 |
| **News** | 118 | 71 |
| **Wikipedia** | 121 | 72 |
| **Blogs** | 124 | 78 |
| **Facebook** | 124 | 76 |

The results are summarized in Figure 3: The results for citations are in agreement with the expectations listed in Table 4. The correlation between citations and REF scores for PRO publications is $r_s$=0.57, 95% CI [0.45, 0.67]. This is close to the result of the meta-analysis summarizing the correlation coefficients from other studies that have correlated REF scores and citation impact ($r_{pooled}$ = 0.71, 95% CI [0.64, 0.78] – see Figure 1). The 95% CIs of both results overlap. The difference between the correlations might be the result of our choice of a relatively small citation window (see section 7.3).

The coefficient for the correlation between REF scores for societal impact and citation impact for PCS publications referenced in case studies is close to zero, $r_s$=0.09, 95% CI [-0.27, 0.11]. This is in agreement with our expectation that case study research references are used in connection with societal impact and not (necessarily) of academic research impact.

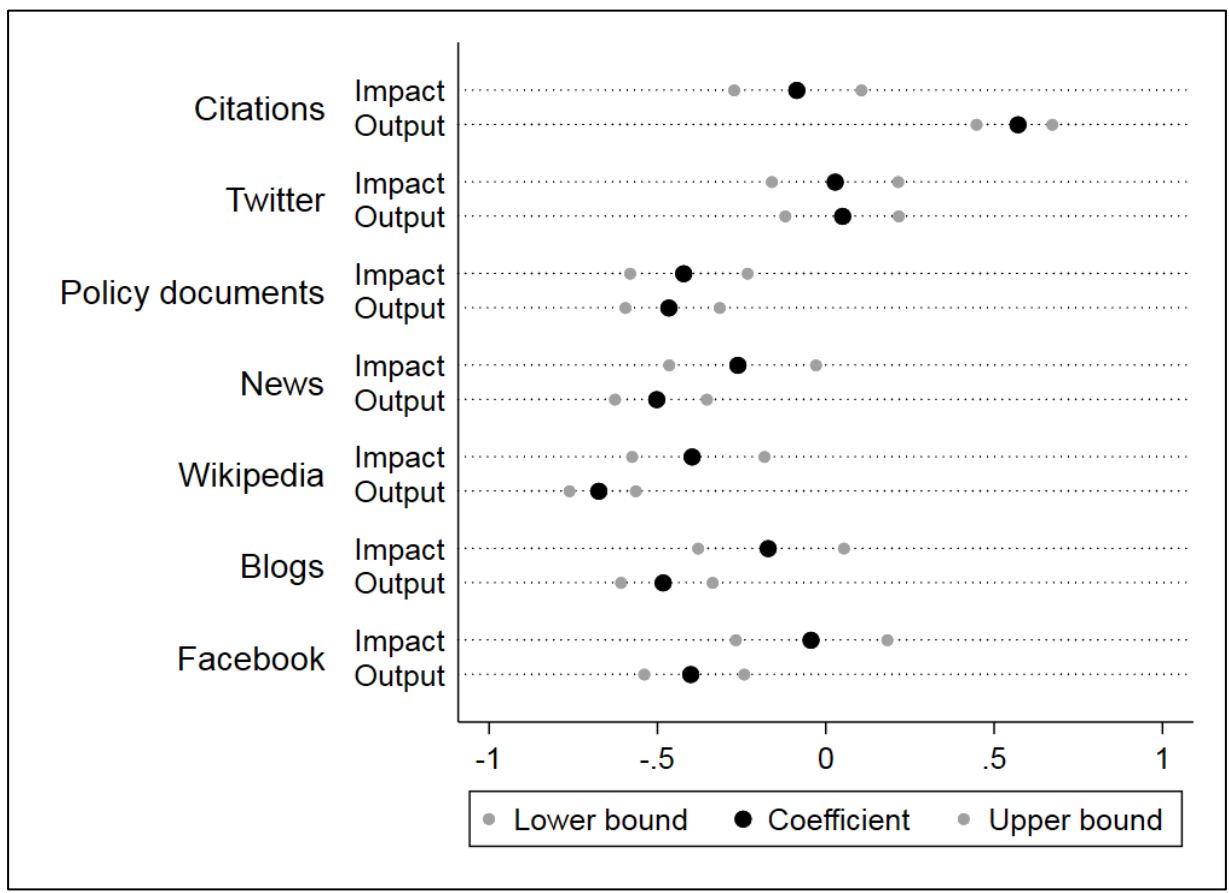

Figure 3. Spearman rank order correlations between MHq' values (citations and altmetrics) and REF scores in the assessment of societal impact and research output. The altmetrics are sorted by the difference between MHq' values for societal impact and research output.

The results in Figure 3 concerning altmetrics do not fully support the expectations in Table 4. Correlations between REF scores for output and altmetrics scores are either close to zero or negative, as expected, but there are similar coefficients for the correlation between REF scores for societal impact and the altmetrics' scores. These may suggest a divergence between the construct of societal impact used by peer reviewers working in the REF evaluation context and that reflected by altmetrics. Certainly, such results provide no support for the use of altmetrics to replace or even inform peer review in impact evaluation.

## 9  Discussion

Research evaluation has hitherto prioritized the analysis of relative excellence and academic impact, largely through bibliometrics. Public policy is now turning towards the socio-economic impact of research as a return on investment. For the assessment of societal outcomes a different range of indicators will be required: altmetrics have been suggested as a candidate but it is not clear what they capture or whether they reflect societal impact. In this study we draw on altmetrics and citation data for the publications included in the UK Research Excellence Framework (REF). We compare these metrics with the grades awarded via peer assessments. We investigated the convergent validity of altmetrics by comparing two REF datasets: publications as personal research outputs (PRO – which are about individual academic achievement); and publications referenced in case studies (PCS – which are intended to reference research with societal impact). We used the MHq' indicator for assessing impact – an indicator that has been introduced for count data with many zeros. The first part of the analysis suggests that a number of altmetric variables have higher MHq' values for PCS than for PRO (see Figure 2) and thus appear to have convergent validity for these data. However, in the second part of the analysis, where altmetrics are correlated with REF scores on PCS, the negative or close to zero correlations (see Figure 3) challenge the notion of such convergent validity.

Traditional citation analysis is about impact on an academic dimension and data are well-suited to the natural science sector. Scholarly citations do not measure whether a paper is useful in other contexts such as commercialization, work, or teaching (Mohammadi & Thelwall, 2014). According to Haustein, Peters, Bar-Ilan, et al. (2014) "publications are used in the development of new technologies, applied in daily work of professionals, support teaching, and have other societal effects" that cannot be measured by citations from research literature. How, then can such impact be addressed as it rises in policy significance? Proponents of altmetrics claim that they offer the possibility of indexing the impact of papers in a wider range of social sectors (see here Haustein, Peters, Sugimoto, et al., 2014; Moed & Halevi, 2015). For example, public engagement or changes in practice related to society might be reflected in altmetrics data (Khazragui & Hudson, 2015). Bornmann and Haunschild (2017) show promising results on the relationship between Mendeley reader counts of specific user groups and impact assessment in the educational sector.

It is the task of scientometrics to investigate whether altmetrics can indeed be used as indicators of broader impact (Thelwall & Kousha, 2015b). "For an altmetric to be taken seriously, empirical evidence of its value is needed in addition to evidence of a reasonable degree of robustness against accidental or malicious spam" (Wilsdon et al., 2015, p. 43). More widely, research organizations and funders are interested in robust data to measure the impact of their research on stakeholder groups (SIAMPI, 2011). This interest is high "due to a lack of alternatives and arguments by academics that research excellence would itself naturally lead to the desired societal benefits" (Thelwall, 2017, p. 2). In the UK REF, societal impact assessments are now a budget-relevant part of evaluation (NISO Alternative Assessment Metrics Project, 2014). By implication, impact assessment in the REF (and other similar national evaluation procedures) is about a social contract: "a relationship through which researchers gain funding in return for contributing to society" (Terama et al., 2017).

Ravenscroft et al. (2017) investigated whether altmetrics might replace reviewers' assessments of societal impact by focusing on references cited in case studies. They correlated the average Altmetric Attention Score for these with the average REF scores concerning societal impact, using the Altmetric API to append the Altmetric Attention Score (a weighted count over a broad range of sources such as tweets and blog mentions) to the case study references. They found close to zero and negative correlation coefficients, suggesting that these scores measured different things: this challenges the convergent validity of altmetrics with indicators measuring societal impact.

We selected a similar approach by comparing case studies with altmetrics. However, we did not study the Altmetric Attention Score but instead used individual sources of altmetrics. References in REF case studies are excellent data for testing the convergent and discriminant validity of altmetrics data for societal impact measurements. In the first part of our study, we compared altmetrics and bibliometrics for two groups of publications: those referenced in impact case studies (PCS); and those submitted as individual research outputs (PRO). Given the reasons why PCS and PRO are deployed as types of evidence (of impact and of excellence), we should expect high altmetrics scores for PCS (convergent validity) and low altmetrics scores for PRO (discriminant validity), with converse expectations for bibliometric indicators.

Our results reveal that citations, news media, mentions on Facebook, blog and Wikipedia references, and citations in policy-related documents do indeed have the expected convergent and discriminant validity (see Figure 2). While the results for Twitter technically conform to expectation, we doubt that the source is valid for societal impact measurement, because the difference between the MHq' values for PCS and PRO is very small.

In the second part of the study, we correlated REF scores and metrics on the basis of UOA units and by institution. We found that REF scores on impact case studies correlated weakly with altmetrics, which (with Ravenscroft et al., 2017) disqualifies arguments in favor of using altmetrics for societal or broader impact measurements. We refer, for comparison, to the general argument behind the use of citation data as a research performance indicator: that there is a positive correlation between citation impact and expert, peer assessments. For example, editorial decisions in peer review are seen to accord with later citation impact of the accepted (or rejected and elsewhere published) manuscripts (Bornmann, 2011). The results in this study affirm that relationship. By contrast, therefore, the missing link that remains between peers' assessments of societal impact and scores based on altmetrics is a challenge to any application of altmetrics in research evaluation (mirroring the conclusions of Thelwall and Kousha (2015b).

Why do impact assessments using altmetrics and case studies produce different outcomes? Societal impact measurements and altmetrics seem to be very different types of indicators. For example, the timescale of altmetrics may comply with usual timescales in current research evaluations (they focus on most recent years); however, the timescale is not appropriate to measure effects of research on society (which needs a long-lasting perspective). Allen, Stanton, Di Pietro, and Moseley (2013) contend – based on a study using articles in the clinical pain sciences – that "the most common altmetrics are not measuring impact, insofar as impact relates to the effect of research on clinical practice or thinking". Providers of altmetrics data, such as the company Altmetric, have already responded to this and other mismatches between altmetrics and requirements of societal impact measurements by toning down the potential of altmetrics data for measuring impact (see Moed, 2017).

A further factor is the time-frame (see section 6) and the emergence of demonstrable causal links. Alla et al. (2017) systematically reviewed the literature for definitions of research impact to develop a general definition of research impact in the area of public health research: "Research impact is a direct or indirect contribution of research processes or outputs that have informed (or resulted in) the development of new (mental) health policy/practices, or revisions of existing (mental) health policy/practices, at various levels of governance (international, national, state, local, organisational, health unit)". Similar definitions of societal impact can be found without the specific focus on public health research (Samuel & Derrick, 2015). These definitions link 'impact' with concrete actions that <u>follow</u> research. The assessed research should make a "change" or "difference" in the world (Samuel & Derrick, 2015). In the context of the REF, Khazragui and Hudson (2015) mention e.g., a spin-out business or a new drug, treatment, or therapy. Whether such actions really took place can be assessed by reviewers (in the REF procedure based on the descriptions in the case studies) and drive the corresponding impact scores: "A positive evaluation outcome will be as dependent on how 'convincing' the case study constructs the strong causal link between the underpinning research and the impacts being claimed" (Derrick, 2014, p. 141).

However, this reflective context is not that in which altmetrics are employed. Users of social platforms do not assess concrete actions but instead mention papers of interest that might also be of utility: in this study, publications referenced in case studies (PCS) receive higher altmetrics scores than publications submitted as research outputs (PRO). Altmetrics based on social media data do not capture whether such research made a "difference" or "change" in the world. The limitation of altmetrics is recognizable in the time horizon of altmetrics: most activity follows immediately after research publication. By contrast, the realization of societal or technological effects has a much greater time lag, as long established by Griliches (1986) and Mansfield (1991). For example, a survey of corporate R&D executives revealed "that an average of 6 years elapsed between a research finding and commercialization" (Khazragui & Hudson, 2015, p. 54).

Reviewers investigating societal impact are interested in the causal link between research and practice/outcome. They assess whether research results have really been substantiated in concrete actions. They use case studies to test what we really want to know: was the research practically useful? The case studies approach has been criticized, especially because the universities only submit success-stories. Research policy might be better informed by capturing the research impact of an institution more holistically, which is one reason why research policy is interested in quantitative indicators of diverse and broader impact. Wilsdon et al. (2015) mention two altmetrics which might serve for societal impact measurement: "only Google patent citations and clinical guideline citations can yet be shown to reflect wider societal impact" (p. 49). Altmetrics based on social media data are much further from measuring realized societal impact, capturing only unknown attention or unstructured noise produced by published research.

For developing robust and valid societal impact metrics, Pollitt et al. (2016) applied the survey-based "best–worst scaling" (BWS) method. The authors asked participants to assess statements about different impact types (e.g., "research helps create new jobs across the UK"). The results of the survey were used "to develop a model that elicited the perceived value of different types of research impact for different groups and segments of survey respondents, including whether the public have different valuations from researchers" (Pollitt et al., 2016). Based on the model, they were able to explore valuations for research impact of different stakeholders. The authors found, e. g., that improved life expectancy is valuable for both the general public and researchers, but differences between both groups exist, e. g., with regards to commercial capacity development. The group-specific valuations for research impact which result from the survey can be used to develop or find metrics which correspond to the valuations. Thus, the study of Pollitt et al. (2016) suggests that impact measurements do not start with available data (as it seems to be the case with altmetrics), but with concrete indications of measurements requirements.

Whereas Pollitt et al. (2016) used the survey-based BWS method to explore valuable and group-specific directions of impact measurements, Hicks, Stahmer, and Smith (2018) propose to use lists of central capabilities (e.g., the set of basic human needs and values by Nussbaum, 2000) and to translate the capabilities into metrics. Whereas the approach by Pollitt et al. (2016) seems to be suitable for developing metrics for specific evaluation contexts (e.g., which focus on certain groups or fields), the approach by Hicks et al. (2018) might be useful in large-scale evaluations such as the UK REF (or similar national evaluation systems) including many units from various disciplines.

# Acknowledgements

We are grateful for the cogent observations and criticisms of the reviewers of an earlier version of this paper. The bibliometric data used in this paper are from a custom database of the Competence Center for Bibliometrics (http://www.bibliometrie.info/). Altmetrics data were used from a locally maintained database using data shared by the company Altmetric on October 02, 2017. The REF output data were downloaded from http://results.ref.ac.uk/DownloadSubmissions/ByForm/REF2 on September 28, 2017. The REF case study IDs and corresponding DOIs were shared with us by Digital Science on December 11, 2017.